# Mutual synchronization of spin torque nano-oscillators through a non-local and tunable electrical coupling

R. Lebrun[1], S. Tsunegi[1,2], P. Bortolotti[1], H. Kubota[2], A.S. Jenkins[1], M. Romera[1], K. Yakushiji[2], A. Fukushima[2], J. Grollier[1], S. Yuasa[2] and V. Cros[1,*]

[1] *Unité Mixte de Physique CNRS, Thales, Univ. Paris-Sud, Université Paris-Saclay, 91767 Palaiseau, France*
[2] *Spintronics Research Center, National Institute of Advanced Industrial Science and Technology (AIST), Tsukuba, Japan*

**Abstract**

The concept of spin torque driven high frequency magnetization dynamics has opened up the field of spintronics to non-linear physics, potentially in complex networks of dynamical systems. In the scarce demonstrations of synchronized spin-torque oscillators, the local nature of the magnetic coupling that is used have largely hampered a good understanding and thus the control of the synchronization process. Here we take advantage of the non-local nature of an electrical coupling to mutually synchronize spin-torque oscillators through their self-emitted microwave currents. The control of the synchronized state is achieved at the nanoscale through two active components of spin transfer torques, but also externally through an electrical delay line. These additional levels of control of the synchronization capability provide new approaches to underlie a large variety of nanoscale collective dynamics in complex networks.

**Introduction**

In 2005, two seminal papers by S. Kaka et al.[1] and F.B. Mancoff et al.[2], have successfully described the synchronization by spin wave coupling between two closely-spaced spin torque oscillators. Owing to the intrinsic local nature of the spin-wave coupling, a persistent goal in the last decade has been to achieve an in-depth understanding in order to control the synchronization process. Thanks to continuous research efforts[3–9], new demonstrations of synchronization through dipolar[10] or spin-wave[11] couplings have been achieved in the recent period. However the alternatives to local magnetic couplings remain to be investigated. The non-local electrical coupling, theoretically proposed in 2006[12], represents such a promising approach, which is anticipated to make spin torque oscillators as a table-top model implementation[13,14] of the underlying physics of non-linear phenomena, e.g. total, partial or chaotic synchronization, in arrays composed of nanoscale dynamical systems[15,16].

Here, we use for the first time the self-emitted radiofrequency current as an efficient source of coupling for achieving the mutual synchronization of two spin torque oscillators and show the predicted improvements in terms of emitted power and spectral coherence. More interestingly, owing to the nature of this coupling mechanism, we provide clear evidence that the different rf-features of the synchronized state (frequency, power, synchronization range and the phase shift between the oscillators) can be finely controlled at the nanoscale through the intrinsic nonlinear parameters of the oscillators and, more originally, through the ratio between the two active components of spin transfer torque, i.e. Slonczewski like (SL) Torque and Field like (FL) Torque. We also show that the coupling can be externally tuned through an electrical delay line. The full control of the synchronization capability represents a specific and definite advantage of spin torque oscillator not only to investigate the complex features of collective nano-scale and nonlinear physics[14–17] but also to mimic basic functionalities of the brain[13,18–20] with oscillator networks.

**Mutual synchronization through self-emitted microwave currents**

The main purpose of this study is to experimentally investigate the electrical coupling between two STOs as a possible source of interaction in order to achieve synchronization. To reach this goal, two types of electrical connection between the STOs might be envisaged, namely either a connection in series or in parallel, even though the physical mechanisms responsible for the synchronization will be the same in both cases[12,21,22]. In the experiments presented here, we will consider the latter case of a parallel connection between the two spin torque oscillators. As shown in Fig. 1, each STO is independently supplied by a dc current source, allowing them to enter in a regime of sustained oscillations through the action of spin torque. The microwave signal emitted by both oscillators is the mechanism responsible for mutual synchronizing them. This is achieved by having the microwave ports of the two bias tees electrically connected through microwave cables and a tunable delay line. Finally, we insert a power splitter (PS) in the circuit in order to record the output microwave signal originating from the two STOs using a spectrum analyzer.

An important condition for the experimental observation of the mutual synchronization between STOs is that the synchronizing force has to be larger than the thermal fluctuations[23]. In addition, we emphasize that the condition to get mutual synchronization requires STOs with a narrow linewidth and a large output power together with an efficient injection locking process to an external microwave current. These considerable requirements motivated our choice to work with vortex based STOs, which show the desired microwave properties. Here we use STOs based on the spin transfer induced dynamics of two interacting vortices in a spin-valve. From our previous studies [24–26], we have demonstrated precisely the vortex configurations and the spin torque components (the ones associated to the vortex-like spin polarization) that can result in a sustained dynamical state at room temperature showing a strongly coherent (~ 100 kHz) and powerful (~ 400 nW) emitted radio frequency signal [25].

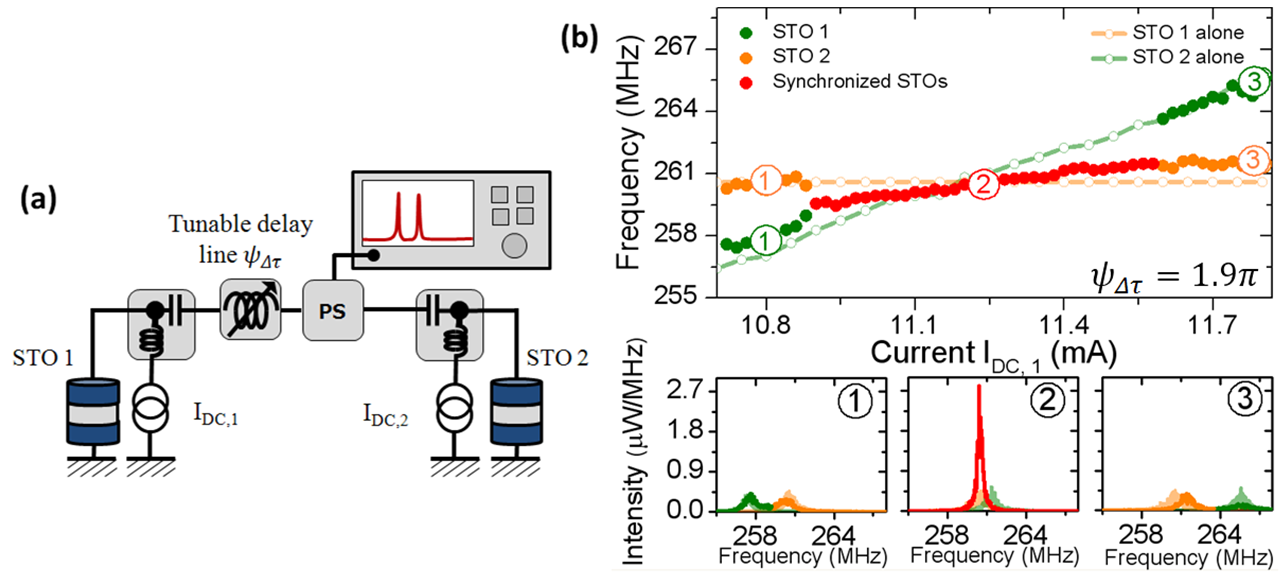

Fig. 1 (a) Scheme of the electrical circuit for the mutual synchronization of two oscillators independently supplied by two currents and connected through the microwave port of two bias tees with a tunable delay line $\psi_{\Delta\tau}$. The detected signal is measured by a spectrum analyzer, connected to the delay with a -6 dBm power splitter (PS). (b) Evolution of the frequency of the interacting STOs as a function of $I_{DC,1}$ while $I_{DC,2}$ is fixed to + 10.6 mA. (Bottom) Corresponding spectra for $I_{DC,1}$= +10.8 mA (1), +11.25 mA (2), +11.8 mA (3) (Non-interacting oscillator properties when one is switched off are in orange and green softened curves).

Our approach relies upon the electrical coupling mechanism (as opposed to local coupling based on magnetic interactions [1,2,5]) and allows us to have access to the dynamical properties of each oscillator when they are interacting or when they are independent. This provides a unique opportunity to properly characterize the microwave properties of the synchronized state by comparing the signal recorded on the spectrum analyzer in two series of measurements. A first set of measurements is recorded when the two STOs are self-oscillating due to the spin transfer torque (the two dc sources supplying the STOs are switched ON, see Fig. 1.a). For that, we keep the applied dc current constant on STO2 and sweep the current applied on STO1. Note that these measurements have been recorded for an optimized electrical delay length $\psi_{\Delta\tau} = 1.9\pi$ (expressed in period of STO's oscillations); the influence associated with a change of the delay will be discussed in the last section of this article. In Fig. 1.b, we display the first experimental demonstration of mutual synchronization between STOs via electrical coupling. Indeed, in region 2 (red dots in Fig. 1.b) we observe a single peak having a much larger power than the two peaks outside the synchronization bandwidth (region 1 and 3 with green and orange dots in Fig. 1.b). This single peak in region 2, where the two STOs have a common frequency, is observed over a frequency range equal to 2 MHz which corresponds to the synchronization bandwidth $\Delta\omega_{sync}$.

In a second set of measurements, we have recorded independently the microwave signal from each STO while the current supplied to the other STO is zero in order to compare quantitatively the emitted signals in the interacting and non-interacting states. Data corresponding to STO1 ($I_{DC,1}$ is ON, $I_{DC,2}$ is OFF) and STO 2 ($I_{DC,2}$ in ON, $I_{DC,1}$ is OFF) measured independently are respectively shown in light green and in light orange in Fig. 1.b. A more complete characterization of the individual STOs can be found in the supplementary information. In the region of mutual synchronization (region 2 on Fig. 1.b), we notice that the emitted signal is much larger than for the non-interacting STOs. The frequency of the synchronized state also differs from the frequency of either of the two non-interacting STOs indicating that the synchronization process is not unidirectional and that the two STOs are mutually synchronized. Furthermore, the spectral coherence in this region 2 is also increased compared to the non-synchronized state as shown in the bottom graphs of Fig. 1b.

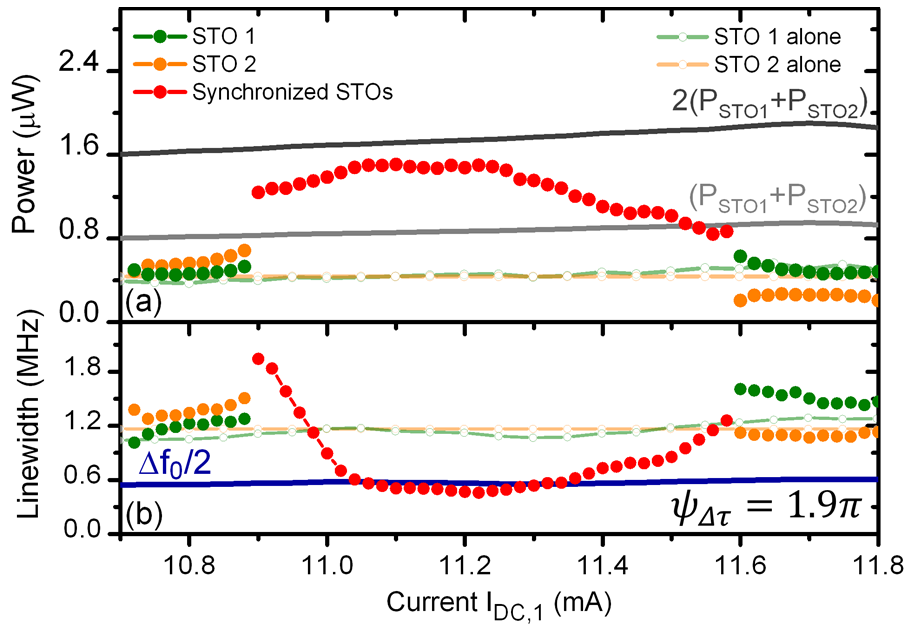

*Fig. 2 Emitted microwave power (a) and linewidth (b) depending on the dc-current injected in the oscillator STO 1 ($I_{DC,2}$ is fixed to + 10.6 mA for STO 2). The characteristics of each STO when one is switched off are shown in green and orange softened curves).*

To analyze more quantitatively the microwave features of the mutually synchronized STOs, we evaluate the emitted power and the spectral coherence in the synchronized state and in the non-interacting states as shown in Fig. 2. First we notice that the two non-synchronized oscillators have very similar power amplitude ($P_{STO1} \sim P_{STO2}$) and linewidths ($\Delta f_{STO1} \sim \Delta f_{STO2}$) (see light green and orange dashed line in Fig. 2.b). This is important because we can thus consider the two STOs as almost identical except that they differ only in frequency. As a consequence, their respective emitted powers and linewidths are close to their average mean value $P_0$ and $\Delta f_0$. As shown in Fig. 2, we observe a maximum of power close to the center of the synchronization bandwidth $\Delta \omega_{sync}$, minima at the edges and two important conclusions can be drawn. First, the measured power in the synchronized regime is superior to $2P_0$, the sum of the two non-interacting emitted STO power. This experimental observation demonstrates *per se* that we do really achieve the mutual synchronization of the two STOs. In the best experimental conditions, i.e., the center of the synchronized regime, we find that the total emitted power $P_{tot}$ reaches almost $4P_0$. We emphasize that such strong and quantitative power enhancement in the synchronized state at zero frequency detuning is theoretically expected but has never been observed until now. This is a crucial advance in order to tackle the important issue of the emitted power of nanoscale STOs[13], which until now has represented an insurmountable roadblock for a range of proposed STO-based applications.

In parallel to the strong increase of the emitted power in the synchronized state, we also find a remarkable improvement of the spectral coherence [1,2,5,6]. As shown in Fig. 2.b, we observe that the linewidth of the synchronized peak at the center of the synchronization bandwidth is reduced down to 550 kHz, corresponding to a reduction by a factor of two compared with the non-synchronized states. Such a significant reduction provides a clear confirmation that the phase noise in the synchronized state is driven by the diffusion of the phase sum[16,23]. Note that at the edges of the synchronization bandwidth, the measured linewidth becomes larger than the ones of the independent STOs which is most probably associated with either frequency pulling and/or phase slips resulting in temporary loss of the synchronized state[10,26]. To our knowledge, our report is the first quantitative confirmation that both the spectral coherence and emitted power in the synchronized state of N-synchronized STOs, whatever the coupling mechanism, can be respectively enhanced by a factor $N$[16,27] and $N^2$[12].

**Control of the synchronization state through a tunable delay**

The measurements shown in Fig. 1 and Fig. 2 have been performed for an optimized delay between the two STOs. Hereafter we present how this delay can be a crucial parameter to control the synchronized state. Indeed, the synchronization bandwidth $\Delta \omega_{sync}$, i.e., the frequency range in which the two STOs have a common frequency, is predicted to depend not only on the strength of the synchronizing force $F_e$, but also on the phase difference between the two STOs[22,28]:

$$\Delta \omega_{sync} = 2F_e \cos(\psi_{\Delta \tau} - \psi_e) . (1)$$

with $\psi_{\Delta \tau}$ the delay introduced by the delay line[29], $\psi_e$ the intrinsic phase shift between the two STOs[26]. Our approach has been to introduce an electrical delay line (see Fig. 1.a) that allows us to tune the total phase difference through the control of the delay constant $\psi_{\Delta \tau}$. The evolution of the mutual synchronization bandwidth as a function of the delay is presented in Fig. 3 in which a $\pi$-periodic oscillation of the synchronization bandwidth with the delay constant is clearly observed. Indeed, by selecting the proper delay $\psi_{\Delta \tau}$, we can either maximize (Fig. 3.d) or minimize (Fig. 3.b) the synchronization bandwidth. We note that the maximum amplitude of the synchronization bandwidth does not exceed 2 MHz only because of the modest amplitude of the synchronizing force $F_e$ given the input parameters (dc current, magnetic field) used for these measurements. Another important result from Fig. 3 is that the maxima (resp. minima) of the synchronizing bandwidth are observed for delays $\psi_{\Delta \tau}$ around $9\pi/10$ (modulo $\pi$) (resp. $2\pi/5$ (modulo $\pi$)). As we will show in the following, this observation is crucial to determine the origin of the synchronizing force.

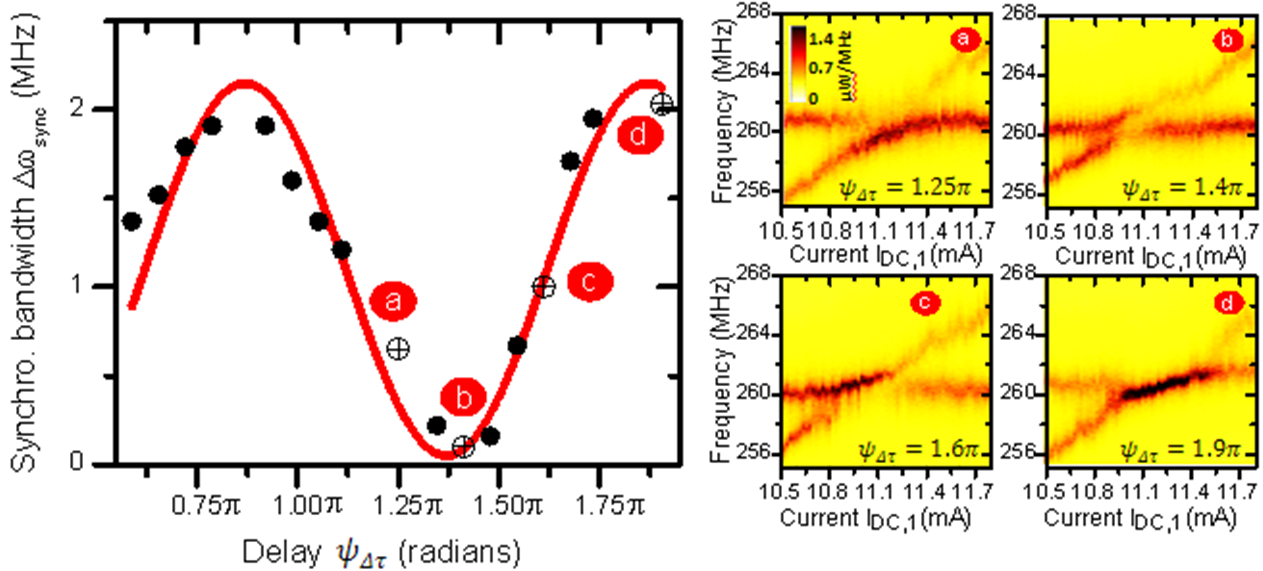

*Fig. 3 Evolution of the synchronization bandwidth of the two mutually synchronized oscillators (STO 1 with $I_{DC,2}$ = +10.6mA and STO 2 with $I_{DC,1}$ swept) depending on the delay constant (associated to the length of the delay line). Right: Color maps of the power spectral density as a function of frequency and $I_{DC,1}$ at different delays: intermediate synchronization bandwidth (a) for $\psi_{\Delta\tau} = 1.25\pi$, minimum (b) for $\psi_{\Delta\tau} = 1.4\pi$, intermediate (c) for $\psi_{\Delta\tau} = 1.6\pi$ and maximum (d) for $\psi_{\Delta\tau} = 1.9\pi$.*

The implementation of a delay line between the two STOs is thus a crucial tool for controlling the synchronization bandwidth. Here we show that it also allows the control of the frequency of the mutually synchronized state $\omega_s$ which, thus, does not depend only on the eigenfrequency of each oscillator but also on the frequency detuning $\Delta\omega = \omega_{STO,2} - \omega_{STO,1}$ between the two STOs [7]:

$$\omega_s = \bar{\omega} - \sqrt{\Delta\omega_{sync}^2 - \Delta\omega^2}\tan(\psi_{\Delta\tau} - \psi_e) . (2)$$

From Eq. 2, we observe that the synchronized frequency $\omega_s$ equals the average frequency $\bar{\omega} = (\omega_{STO,2} + \omega_{STO,1})/2$ (with $\omega_{STO,2}$ and $\omega_{STO,1}$ the frequencies of the non-interacting STOs) when the phase difference $\psi_{\Delta\tau} - \psi_e$ is close to zero (mod $\pi$), a condition that is fulfilled when the synchronization bandwidth is maximum (see Fig. 3). As shown in Fig. 3d, the predicted synchronized frequency is in agreement with the experimental frequency of the synchronized state. For an intermediate synchronization bandwidth, i.e. $|\psi_{\Delta\tau} - \psi_e| \sim \pi/4$, the frequency in the synchronized regime differs from $\bar{\omega}$, the average frequency. Depending on the sign of the tangent term in Eq. 2, the frequency can be either larger or smaller than $\bar{\omega}$. This is confirmed experimentally as displayed in Fig. 3.a and Fig. 3.c where the synchronized frequency is either higher (for $\psi_{\Delta\tau} \sim 1.6\pi$ rad) or lower (for $\psi_{\Delta\tau} \sim 1.25\pi$ rad) than the average frequency for intermediate synchronization bandwidths. Note that our analysis is performed in the middle of the synchronization bandwidth ($\Delta\omega \sim 0$) where there is no frequency pulling. The microwave features of two synchronized STOs thus strongly depends on the delay between them, which could be of a great interest for tuning the filtering functionality of arrays of synchronized STOs[20] or in bio-inspired associative memories[13,30,31].

**Field-like torque as a mean to drive the intrinsic phase shift between synchronized oscillators**

From these aforementioned observations we can deduce that a maximum synchronization bandwidth is expected for zero electrical delay between the two STOs. Such a feature raises the prospect of the synchronization of multiple STOs without the necessity to add a large length of microwave cables between each oscillator and so to avoid detrimental constraints for real applications[32]. Indeed, the synchronization bandwidth (see Eq. 2) of two STOs depends on their intrinsic phase shift $\psi_e$, which is expressed for a vortex based STO as:

$$\psi_e = \tan^{-1} \nu + \tan^{-1}\left(\frac{\Lambda_{FL}}{\Lambda_{ST}}\right). \quad (3)$$

We emphasize that this phase shift is not only related to the term associated to the nonlinear oscillator parameter $\nu$ [22] (as it is often the case for other types of oscillators[7,14,33]) but also to the ratio of the two components of spin torque responsible for the synchronization, i.e. the Slonczewski like Torque ($\Lambda_{SL}$) and the Field like Torque ($\Lambda_{FL}$)[23,34]. Through the analysis of time domain measurements, we can experimentally determine the nonlinear parameter ν [23] resulting in $\tan^{-1}\nu \sim 2\pi/5$ in Eq. 3. Moreover, we know from our previous studies on vortex based STOs [26,35] that the force associated to the Field like torque is large compared to the Slonczweski like Torque leading to $\tan^{-1}\left(\frac{\Lambda_{FL}}{\Lambda_{ST}}\right) \sim \pi/2 \pmod{\pi}$. Thus, maxima of the synchronization bandwidth are expected for $\psi_{\Delta\tau}$ close to $9\pi/10 \pmod{\pi}$ which is in excellent agreement with the results represented in Fig. 3. This robust observation indicates that a large synchronization bandwidth $\Delta\omega_{sync}$ can be obtained for $\psi_{\Delta\tau} = 0$, i.e., without delay line. This remarkable feature also highlights the crucial role of the Field like Torque in the synchronization of these vortex STOs, compared to the case of uniform based STOs, for which all the pioneering theoretical studies have considered the Field like torque to be negligible in the synchronization process[7,32,33]. Interestingly, one can notice that the efficiency of the different locking torques is now known to change as a function of the bias voltage[36,37] which provides an additional parameter to optimize the intrinsic phase shift. The full control of the electrical synchronization of two STOs, both externally with an electronic delay and intrinsically through the ratio of the spin transfer torques, marks an important milestone towards the observation of large variety of nanoscale collective dynamics of nonlinear oscillators[15], and opens among others the perspective of STO networks mimicking some of the basic brain functionalities[13].

## Methods

The spin transfer oscillators that have been investigated are circular-shape magnetic tunnel junctions (MTJ) with a nominal diameter of 300 nm. The multilayer stacking of each MTJ is composed of a double vortex spin-valve on top of a magnetic tunnel junction: Synthetic antiferromagnet (SAF) / MgO (1.075) / NiFe (6) / Cu(9.5) / NiFe (20) (with thickness in nm). The pinned SAF layer is a PtMn (15) / CoFe (2.5) / Ru (0.85) / CoFeB (3) multilayer. The two NiFe layers have a magnetic vortex as ground state. The GMR ratio of the Cu based spin-valve is about 2% whereas the TMR ratio of the MgO based MTJ is about 70% at room temperature and low bias. As a consequence, the output power that is detected on the spectrum analyzer in predominantly arising with the vortex dynamics in the thin NiFe layer that is close to the MgO barrier. More detail about these double vortex based STOs can be found in Ref[25].

The two STOs present in the circuit are sourced individually with two independent dc sources (see Fig. 1). In order to study the electrical mutual synchronization, the two STOs are connected using conventional microwave cables and bias tees. Moreover, in order to tune (manually) the delay time between the two STOs, we have also introduced an electrical delay line as shown in Fig 1, that allows us to vary the the delay time. The detection of the total emitted signal is obtained using a spectrum analyzer connected to the electrical circuit through a -6 dBm power splitter.

In order to prepare magnetically the state in the two STOs, we perform a field cycling before the electrical measurements to initialize each double vortex system in parallel chiralities and antiparallel cores polarities, which permits to have sustained oscillations without any applied magnetic field. In fact, we first apply a large perpendicular magnetic field to impose parallel core polarities and then the field is reversed until the core polarity of the thinner NiFe layer eventually reverses. In our convention, a positive current corresponds to electrons flowing from the thin to the thick NiFe layer, resulting in the spin transfer dynamics of the coupled vortex mode that is mainly located in the thin NiFe layer. See Ref[25] for detail about the characteristic of the different coupled modes and the symmetry of the spin transfer forces.

## Acknowledgements:

The authors acknowledge ANR agency (SPINNOVA ANR-11-NANO-0016) and EU FP7 grant (MOSAIC No. ICT-FP7-8.317950) for financial support.